\documentclass[osajnl,twocolumn,showpacs,superscriptaddress,10pt]{revtex4-1} 
\usepackage{amsmath,amssymb,graphicx}

\usepackage{mathtools}
\usepackage{xfrac}

\begin{document}

\title{On-chip self-referencing using integrated lithium niobate waveguides}

\author{Yoshitomo Okawachi}\email{Corresponding author: y.okawachi@columbia.edu}
\affiliation{Department of Applied Physics and Applied Mathematics, Columbia University, New York, NY 10027}

\author{Mengjie Yu}
\affiliation{John A. Paulson School of Engineering and Applied Sciences, Harvard University, Cambridge, MA 02138}

\author{Boris Desiatov}
\affiliation{John A. Paulson School of Engineering and Applied Sciences, Harvard University, Cambridge, MA 02138}

\author{Bok Young Kim}
\affiliation{Department of Applied Physics and Applied Mathematics, Columbia University, New York, NY 10027}

\author{Tobias Hansson}
\affiliation{Department of Physics, Chemistry and Biology, Link\"{o}ping University, SE-581 83 Link\"{o}ping, Sweden}

\author{Marko~Lon\v{c}ar}
\affiliation{John A. Paulson School of Engineering and Applied Sciences, Harvard University, Cambridge, MA 02138}

\author{Alexander L. Gaeta}
\affiliation{Department of Applied Physics and Applied Mathematics, Columbia University, New York, NY 10027}
\affiliation{Department of Electrical Engineering, Columbia University, New York, NY 10027}

\begin{abstract}
The measurement and stabilization of the carrier-envelope offset frequency $f\textsubscript{CEO}$ via self-referencing is paramount for optical frequency comb generation which has revolutionized precision frequency metrology, spectroscopy, and optical clocks. Over the past decade, the development of chip-scale platforms has enabled compact integrated waveguides for supercontinuum generation. However, there is a critical need for an on-chip self-referencing system that is adaptive to different pump wavelengths, requires low pulse energy, and does not require complicated processing. Here, we demonstrate efficient carrier-envelope offset frequency $f\textsubscript{CEO}$ stabilization of a modelocked laser with only 107 pJ of pulse energy via self-referencing in an integrated lithium niobate waveguide. We realize an $f$-$2f$ interferometer through second-harmonic generation and subsequent supercontinuum generation in a single dispersion-engineered waveguide with a stabilization performance equivalent to a conventional off-chip module. The $f\textsubscript{CEO}$ beatnote is measured over a pump wavelength range of 70 nm. We theoretically investigate our system using a single nonlinear envelope equation with contributions from both second- and third-order nonlinearities. Our modeling reveals rich ultrabroadband nonlinear dynamics and confirms that the initial second harmonic generation followed by supercontinuum generation with the remaining pump is responsible for the generation of a strong $f\textsubscript{CEO}$ signal as compared to a traditional $f$-$2f$ interferometer. Our technology provides a highly-simplified system that is robust, low cost, and adaptable for precision metrology for use outside a research laboratory.
\end{abstract}


\maketitle 


\section{Introduction}
\label{sec:intro}

\begin{figure}[!b]
\centering
  \includegraphics*[width=\linewidth]{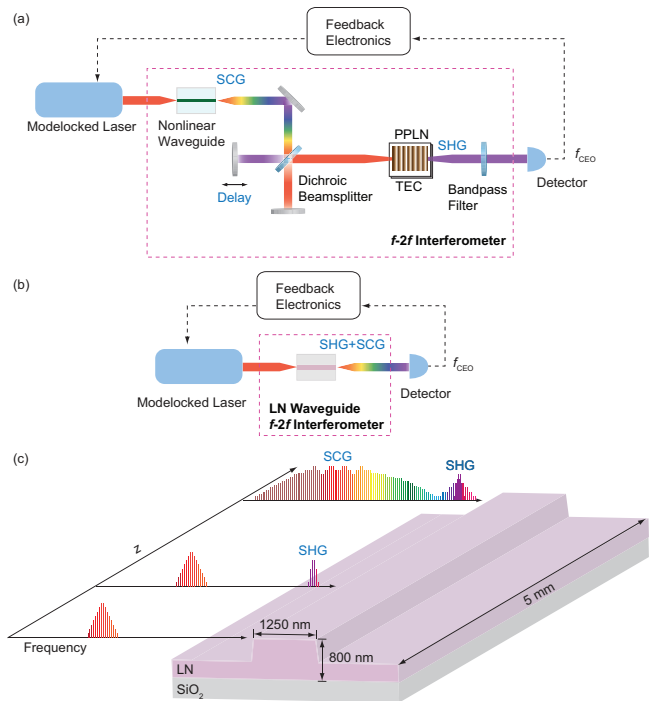}
  \caption{(a) Conventional $f\textsubscript{CEO}$ detection using nonlinear waveguide for SCG and a free-space $f$-$2f$ interferometer with periodically-poled lithium niobate (PPLN) for SHG. (b) $f\textsubscript{CEO}$ detection using integrated lithium niobate (LN) waveguide. (c) Illustration of spectral evolution in LN waveguide. TEC: thermoelectric cooler.}
  \label{Fig1}
\end{figure}

The development of optical frequency combs has enabled high-precision frequency measurements and led to advances in a wide area of research including all-optical clocks, spectroscopy, and metrology \cite{udem02,Cundiff03,Diddams10}. Significant advances in nanofabrication technology over the past decade have led to the development of various chip-based platforms for frequency comb generation, including silicon nitride, silicon dioxide, silicon, and aluminum nitride \cite{Gaeta19,Duchesne10,Kuyken11,Halir12,Kuyken2015,Singh15,Liu16,Klenner16,Carlson17,Porcel17,Okawachi18,Waldburger19,Hickstein19,Oh17,Singh18,Hickstein17,Pfeiffer17,Brasch17,Drake19,Newman19}. Over the past two decades, two different approaches have been developed for on-chip frequency comb generation. One approach is based on stabilization of the repetition rate and carrier-envelope offset frequency ($f\textsubscript{CEO}$) of a modelocked laser. The $f\textsubscript{CEO}$ can be detected using a self-referenced $f$-2$f$ interferometer, which requires a phase coherent octave-spanning spectrum  [Fig. \ref{Fig1}(a)] \cite{Telle99,Jones00,Holzwarth00,Cundiff03}. This broadband spectrum is achieved through supercontinuum generation (SCG) in a nonlinear waveguide. The second approach involves Kerr comb generation (KCG), where a single-frequency, continuous wave laser is used to pump a high-$Q$ microresonator to excite a broadband, dissipative Kerr soliton through parametric four-wave mixing \cite{Gaeta19}. While the nonlinear broadening stage has been implemented on-chip, $f$-$2f$ interferometry has been largely performed using bulk optics and a periodically poled $\chi^{(2)}$ crystal or waveguide for second harmonic generation (SHG) \cite{Jones00,Holzwarth00,Klenner16,Carlson17,Waldburger19}. Since this process occurs after spectral broadening though SCG or KCG, the spectral components used for harmonic generation are at the wings of the generated spectrum, limiting the available peak power and resulting in low power conversion efficiency of SHG. This issue is particularly severe in KCG \cite{Brasch17,Drake19,Newman19}, where auxiliary lasers locked to the Kerr comb are frequency doubled or tripled for $f$-2$f$ or 2$f$-3$f$ interferometry. In addition, a variable delay line needs to be implemented in such a system to compensate for the dispersive walk-off between the $f$- and $2f$-components. Furthermore, for efficient phase matching at different wavelengths, devices with different poling periods are needed and precise temperature control is required.

\begin{figure*}[t]%
\centering
  \includegraphics*[width=\textwidth]{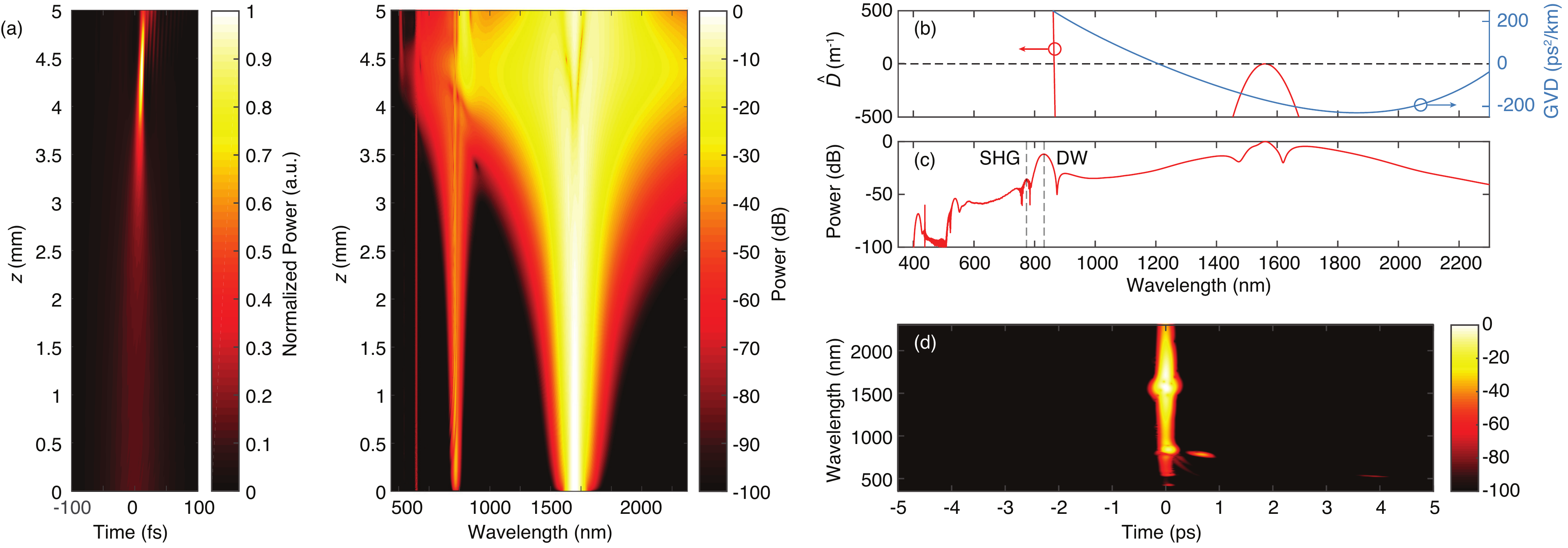}
  \caption{%
 Numerical simulations of pulse propagation in a LN waveguide. (a) Simulated temporal (left) and spectral (right) evolution in a 0.5-cm-long LN waveguide with a cross section of 800$\times$1250 nm. (b) Simulated group-velocity dispersion (GVD, blue) and calculated dispersion parameter ($\hat{D}$, red) and (c) simulated output spectrum. Waveguide cross section is 800$\times$1250 nm. The spectrum shows generation of second and third harmonic components (labeled SHG and THG, respectively) along with a dispersive wave (DW). (d) Simulated spectrogram at the waveguide output.}
  \label{Fig2}
\end{figure*}

As an alternative, here we consider a scheme where high peak power pump pulses first generate a harmonic signal before the remaining pump is used for SCG to create the fundamental frequency component [Fig. \ref{Fig1}(b),(c)]. The high  peak power allows for highly efficient $\chi^{(2)}$-based harmonic generation, while providing sufficient excess pump power to allow for spectral broadening through the $\chi^{(3)}$ nonlinear process. Recently, there have been demonstrations of on-chip $f$-$2f$ interferometry through simultaneous SCG and second-harmonic generation (SHG) \cite{Okawachi18,Hickstein19,Hickstein17}, and $f\textsubscript{CEO}$ stabilization has been demonstrated in silicon nitride (SiN) waveguides using a photo-induced nonlinear grating effect ($\chi^{(2)}$=0.5 pm/V) \cite{Hickstein19} and aluminum nitride waveguides ($\chi^{(2)}$=1 pm/V)\cite{Hickstein17}. While such an approach offers the potential for a high level of simplicity to produce a self-referenced frequency comb, SiN waveguides require an optical-writing process, which involves a femtosecond laser to generate the effective  $\chi^{(2)}$ nonlinearity and sets a limit on the input pulse energy that can be used for $f$-$2f$ interferometry, and AlN waveguides demands nanojoule pulse energies which is considerably higher than what has been achieved with separate SCG and SHG systems \cite{Klenner16,Carlson17,Waldburger19}.

In recent years, integrated lithium niobate (LN, LiNbO$_3$) has emerged as an ideal platform for nonlinear photonics and its large nonlinear index ($n_2 = 2\times 10^{-19}$ m$^2$/W) and strong $\chi^{(2)}$ nonlinearity ($\chi^{(2)}$= 40 pm/V) \cite{WangComm,WangOptica,Yu19,Lu,Shoji97}. Moreover, advances in waveguide fabrication technology \cite{ZhangOptica} has led to the realization of low-loss waveguides with tight optical confinement, enabling dispersion engineering which is critical for nonlinear photonics applications. Previously, Yu, \emph{et al.} \cite{Yu19} has shown the first evidence of $f\textsubscript{CEO}$ detection using octave-spanning SCG in a LN waveguide. Alternatively, SCG has been demonstrated in a periodically poled integrated LN waveguide via cascaded nonlinearities using a 2-\textmu m pump \cite{Jankowski20}. However this system produces a weak $f\textsubscript{CEO}$ beatnote due to the low pulse energy, and requires further complexity in design in terms of both dispersion engineering and group velocity matching and fabrication for poling.

In this paper, we demonstrate highly efficient self-referencing in an integrated LN waveguide by leveraging the large intrinsic $\chi^{(2)}$ and $\chi^{(3)}$ nonlinearities. Self-referencing is achieved by performing both SHG and SCG for $f$-$2f$ interferometry in a single waveguide. We use this LN $f$-$2f$ interferometer to demonstrate $f\textsubscript{CEO}$ stabilization of a modelocked fiber laser with record low pulse energies of 107 pJ, with a large reduction in phase noise $>$100 dB/Hz at 10 Hz. We verify that the  stabilization performance is equivalent to a conventional $f$-$2f$ module. In addition, we demonstrate $f\textsubscript{CEO}$ beatnote detection over 70 nm of pump wavelength tuning. We also numerically model the pulse propagation by employing a single nonlinear envelope equation that incorporates both second- and third order nonlinearities. Our modeling unveils the fascinating underlying dynamics of simultaneous harmonic generation and SCG which manifests in our system and correctly reproduces the experimentally observed spectrum over the vast optical bandwidth spanning multiple octaves. Our demonstration illustrates the technological readiness of LN waveguides for implementation of a low cost and adaptable precision metrology system for use outside a research laboratory.

\section{Theory}
\label{sec:the}

Most of the prior work done on pulse propagation dynamics with $\chi^{(2)}$ effects has implemented coupled equations for the fundamental and second-harmonic fields \cite{Kanashov,Kinsler03,Jankowski20}. However, this analysis breaks down for ultrabroadband $\chi^{(2)}$ and $\chi^{(3)}$ interactions where  these fields spectrally overlap. In order to model ultrabroadband nonlinear phenomena in LN waveguides, where the combined $\chi^{(2)}$ and $\chi^{(3)}$ effects result in multi-octave bandwidth generation, we consider a single nonlinear envelope equation taking into account $\chi^{(2)}$ and $\chi^{(3)}$ effects \cite{Genty07,Conforti,Wabnitz10,Baronio12,Hansson16,Bache16}. We solve the nonlinear envelope equation,

\begin{multline}
  \label{eq:NEE}
 \left[\frac{\partial}{\partial z} - i\sum_{n \ge 2} \frac{\beta_n}{n!} \left(i \frac{\partial}{\partial t}\right)^n +\frac{\alpha}{2} \right]E(z,t)\\
 =i \frac{\omega_0}{2n_0c\epsilon_0}\left(1+i\tau_{sh}\frac{\partial}{\partial\tau}\right)P_{NL}(z,t),
\end{multline}

\noindent where $P_{NL}=\epsilon_0 \left[\chi^{(2)}E^2+\chi^{(3)}E^3\right]$ is the total nonlinear polarization with contributions only from non-negative frequencies, $\tau_{sh}=1/\omega_0-{\partial[\text{ln}(n(\omega))]/\partial\omega}|_{\omega=\omega_0}$ is the optical shock time, $\beta_n$ is the $\it{n}$-th order dispersion coefficient, $\alpha$ is the propagation loss, $\omega_0$ is the pump frequency, and $\tau=t-\beta_1z$ is the local time in the moving frame. We incorporate the effects of second- and third-order nonlinearities, high-order dispersion, and self-steepening. We solve Eq.\,\eqref{eq:NEE} numerically via the split-step Fourier method using the fourth-order Runge-Kutta for the nonlinear step. Figure \ref{Fig2}(a) shows the temporal and spectral evolution of the pulse in a 0.5-cm-long LN waveguide with a cross section of 800$\times$1250 nm. The pump pulse is 90-fs in duration with a pulse energy of 107 pJ and is centered at 1560 nm. In the spectral domain, we immediately see the effects of SHG and sum-frequency generation at 780 nm, along with third-harmonic generation at 520 nm. As the pulse propagates in the waveguide, we observe spectral broadening due to self-phase modulation. For $z>4$ cm, we observe dispersive wave (DW) formation \cite{Efimov05,Dudley,Okawachi17,Webb13} originating near 860 nm that subsequently blue shifts due to phase matching and approaches the second harmonic wavelength. In addition, we observe the formation of the second-harmonic of the DW. Figure \ref{Fig2}(b) shows the simulated group-velocity dispersion (GVD) and the dispersion operator $\hat{D} = \sum_{n \ge 2}\frac{\beta_n(\omega_0)}{n!}(\omega-\omega_0)^n$ for a 1560 nm pump ($\omega_0$ corresponds to the center frequency of the pump) \cite{Dudley,Okawachi17,Webb13}, and Fig. \ref{Fig2}(b)  shows the simulated spectrum at the waveguide output. The spectral position of the DW is predicted from the zero-crossing of the dispersion operator. The spectral overlap between the DW and the second harmonic component allows for effective mixing between the $f$ and $2f$ components and results in a strong $f\textsubscript{CEO}$ beatnote. Figure \ref{Fig2}(d) shows the spectrogram at the output. We calculate a group-velocity mismatch of 130 fs/mm between the pump and the second harmonic component which is significantly lower than that of bulk LN (300 fs/mm) \cite{WangOptica,maddaloni2016laser}. This low temporal walk-off eliminates the need for the implementation of a delay line and enables the single waveguide device for $f$-$2f$ interferometry.  

\begin{figure}[t]
\centering
  \includegraphics*[width=\linewidth]{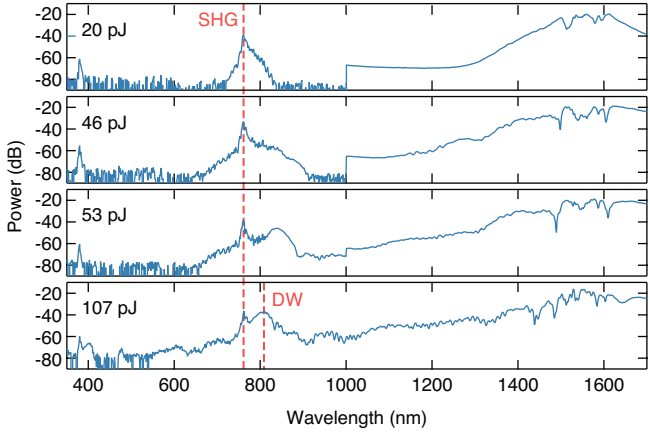}
  \caption{Spectral evolution as the pulse energy in the waveguide is increased (top to bottom). The SHG wavelength is indicated with a dashed (red) line. The discontinuity at 1 \textmu m corresponds to detector switching in the optical spectrum analyzer.}
  \label{Fig3}
\end{figure}

\begin{figure}[t]
\centering
  \includegraphics*[width=\linewidth]{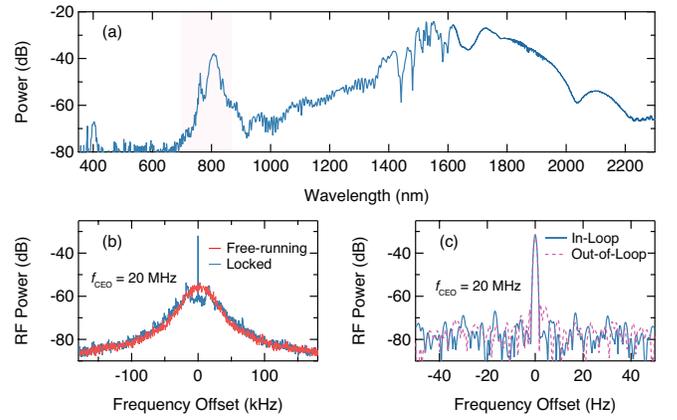}
  \caption{(a) Measured supercontinuum spectrum. Shaded region used for $f$-2$f$ interferometry. (b) Measured RF spectra for free-running (red) and locked $f\textsubscript{CEO}$ beat note (blue). The resolution bandwidth (RBW) is 10 Hz. (c) In-loop (blue) and out-of-loop (green) $f\textsubscript{CEO}$ beat measured over 50 Hz span with 1-Hz RBW.}
  \label{Fig4}
\end{figure}

In our experiment, we send a pulse train from a modelocked erbium fiber laser centered at 1560 nm with a pulse duration of 90 fs and a 250-MHz repetition rate into a 0.5-cm-long air-clad LN waveguide. We pump the fundamental TE mode of the waveguide which allows us to exploit the largest nonlinear tensor component for the $\chi^{(2)}$ process in the x-cut film. The LN waveguide is fabricated using an x-cut 800-nm LN thin film with an etch depth of 450 nm and a width of 1250 nm. The waveguide output is collected using a lensed fiber sent to two different optical spectrum analyzers for spectral characterization. We estimate an input coupling loss of 10.3 dB and an overall insertion loss of 17.5 dB. This coupling loss can be reduced to 1.7 dB, and the overall losses can be as low as 3.4 dB \cite{He19}. Figure \ref{Fig3} shows the spectral evolution as the pulse energy in the waveguide is increased. For 20 pJ of pulse energy, we observe a strong SHG signal peaked at 760 nm and a weak fourth-harmonic signal at 380 nm. As the pulse energy is increased, we observe the formation of a DW centered at 840 nm. At 107 pJ, we observe a blue-shift of the DW due to phase matching that results in overlap with the SHG signal, enabling the generation of a strong $f\textsubscript{CEO}$ beatnote. Similar to our modeling, we observe the second harmonic of the DW near 400 nm. Figure \ref{Fig4}(a) shows the entire supercontinuum spectrum which continuously spans 700 -- 2200 nm for a coupled pulse energy of 107 pJ.

\begin{figure}
\centering
  \includegraphics[width=0.9\linewidth]{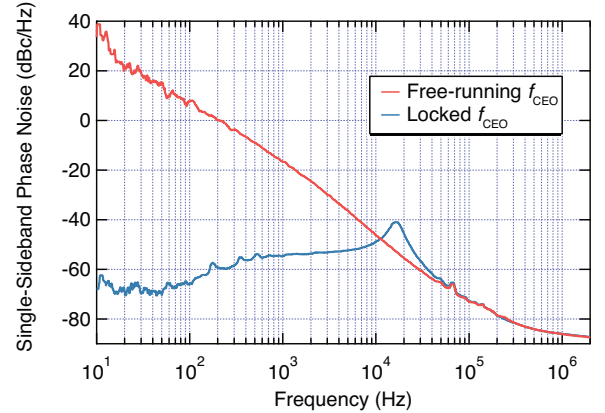}
  \caption{Single-sideband phase noise of the $f\textsubscript{CEO}$ for free-running (red) and locked conditions (blue).}
  \label{Fig5}
\end{figure}

The $f\textsubscript{CEO}$ of the modelocked laser is measured by directly detecting the waveguide output using a silicon avalanche photodiode (APD, 400 -- 1000 nm wavelength range). For $f\textsubscript{CEO}$ stabilization, the measured offset from the APD is phase locked to a 10-MHz rubidium frequency standard using a feedback loop, which includes a phase detector and a PID controller. Figure \ref{Fig4}(b) shows both the measured free-running (red) and locked (blue) in-loop $f\textsubscript{CEO}$ beatnote centered at 20 MHz, obtained with a 10-Hz resolution bandwidth (RBW) using a phase noise analyzer. Figure \ref{Fig4}(c) shows the locked ($f\textsubscript{CEO}$) beatnote over a 50-Hz span with 1-Hz RBW. We measure a 3-dB bandwidth of 1 Hz which is at the resolution limit of the analyzer. For comparison, we measure the out-of-loop $f\textsubscript{CEO}$ beat using a standard $f$-2$f$ interferometer based on a highly nonlinear fiber and a bulk periodically poled lithium niobate frequency doubler, and we observe a nearly identical signature [Fig. \ref{Fig4}(c)]. Figure \ref{Fig5} shows the single sideband phase noise of the $f\textsubscript{CEO}$ beatnote for the free-running (red) and locked (blue) cases. We achieve a tight phase lock and observe a large reduction in phase noise ($>$100 dB/Hz at 10 Hz).

\begin{figure}[!t]
\centering{\includegraphics[width=\linewidth]{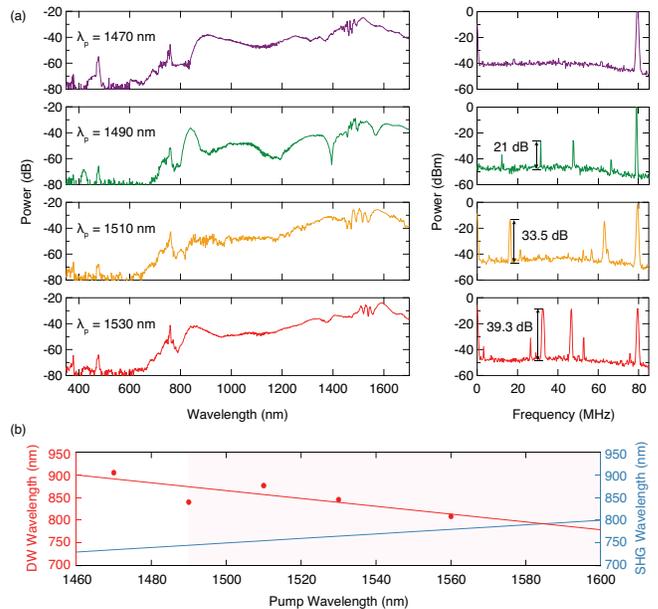}}
\caption{(a) Pump wavelength dependence of $f\textsubscript{CEO}$ beatnote. The plot shows the optical spectra (left) and corresponding RF measurement (right) for four different pump wavelengths from 1470 nm to 1530 nm. The peak at 80 MHz corresponds to the repetition rate and the two highest peaks below that corresponds to $f\textsubscript{CEO1}$ and $f\textsubscript{CEO2}$. We observe a 40 dB increase in the signal-to-noise ratio of the $f\textsubscript{CEO}$ beatnote as the pump wavelength is increased which is attributed to the increased spectral overlap between the DW and the second harmonic component. The difference in the $f\textsubscript{CEO}$ arises from the drift of our pump source. The RBW of the RF spectrum analyzer is 300 kHz. (b) Center wavelength of DW peak (red) and second harmonic of the pump wavelength (blue) for a range of pump wavelengths. Red circles denote the experimentally measured points and the solid red line is a fit based on the points. The shaded region shows the spectra range of $f\textsubscript{CEO}$ detection. We expect the range extends nearly symmetrically on the other side of the crossing point between DW and SHG.}
\label{Fig6}
\end{figure}

Lastly, we investigate the operational range of the pump wavelength for generating the $f\textsubscript{CEO}$ beatnote. For this measurement, we use 200-fs pulses from a tunable femtosecond optical parametric oscillator (OPO) with a repetition rate of 80 MHz. Figure \ref{Fig6}(a) show the measured optical spectra with the corresponding RF spectra. The peak at 80 MHz corresponds to the repetition rate and the two next highest peaks correspond to $f\textsubscript{CEO1}$ and $f\textsubscript{CEO2}$. The pump wavelength is tuned from 1470 nm to 1530 nm, the upper wavelength limited by the operating range of the OPO. We achieved a $f\textsubscript{CEO}$  signal with $>$20 dB signal-to-noise ratio (SNR) with a modelocked pulse source from 1490 nm to 1530 nm. As the pump wavelength is increased, we see the $f\textsubscript{CEO}$ beatnote become stronger with a SNR as high as 40-dB for a pump wavelength of 1530 nm. Remarkably, the $f\textsubscript{CEO}$ beatnotes are bright featuring a high intensity of -8.26 dBm at the same level as the repetition frequency, thanks to the spectral brightness of both DW and SHG components and their relatively good spectral overlap. Since the SHG signal strength largely remains the same, the increase in SNR as the pump is red-shifted is attributed to the blue-shift of the DW towards the second-harmonic position. For this GVD profile (Fig. \ref{Fig2}), the spectral position of the dispersive wave $\omega\textsubscript{DW}$ is largely dictated by GVD and third-order dispersion through the relation $\omega\textsubscript{DW}=-3\beta_2/\beta_3$ \cite{Webb13}. As the pump is red-shifted, the SHG also red shifts while the DW blue shifts due to a increased $\beta_2$. In our waveguide, better spectral overlap between SHG and DW is achieved as the pump wavelength is increased, and an $f\textsubscript{CEO}$ signal $>$20 dB is achieved from 1490 nm to 1560nm. The upper limit is dictated by the tuning range of our pulse source. As we can see from Fig. \ref{Fig3} pumping at 1560 nm, the DW has not yet reached the best overlap with SHG. Figure \ref{Fig6}(b) shows a plot of the peak wavelength of the DW (red) and the second harmonic of the pump wavelength (blue) for a range of pump wavelengths. Based on our fit, we expect the best overlap to occur at 1587 nm, which corresponds to the crossing point between the DW and second-harmonic curves, and we extrapolate that the $f\textsubscript{CEO}$ detection range is nearly symmetric about this crossing point up to 1700 nm. 

In conclusion, we demonstrate on-chip self-referencing using a single integrated LN waveguide. We achieve efficient $f\textsubscript{CEO}$ stabilization of a modelocked fiber laser using 107 pJ of pulse energy by exploiting the efficient second-harmonic process that occurs at the beginning of the waveguide while still allowing for strong $\chi^{(3)}$ interactions with high peak pump power. The platform offers a wide pump wavelength range $>$70 nm over which the $f\textsubscript{CEO}$ beatnote can be generated. The simple structure can replace a conventional $f$-$2f$ interferometer with bulk PPLN which requires various poling periods, a temperature controller, and a delay line for extracting the $f\textsubscript{CEO}$. In addition, we theoretically investigate this system by modeling pulse propagation in a LN waveguide with $\chi^{(2)}$ and $\chi^{(3)}$ effects. The low power consumption and compact footprint of our scheme offers promise towards the miniaturization of frequency comb technology and a step towards the realization of an integrated fully-stabilized frequency comb source for applications beyond the lab.

\section*{Funding Information}
National Science Foundation (NSF) (IIP-1827720, ECCS-1740296 E2CDA); Defense Advanced Research Projects Agency (DARPA) (W31P4Q-15-1-0013); Air Force Office of Scientific Research (AFOSR) (FA9550-15-1-0303, FA9550-19-1-0310, FA9550-19-1-0376).

\section*{Acknowledgments}

Device fabrication is performed at the Harvard University Center for Nanoscale Systems (CNS), a member of the National Nanotechnology Coordinated Infrastructure Network (NNCI), which is supported by the National Science Foundation under NSF ECCS award no.1541959. The authors thank J. K. Jang and Y. Zhao for useful discussions.

\section*{Disclosures}

\medskip

\noindent\textbf{Disclosures.} The authors declare no conflicts of interest.



\bibliography{f2f}

\begin{thebibliography}{47}%
\makeatletter
\providecommand \@ifxundefined [1]{%
 \@ifx{#1\undefined}
}%
\providecommand \@ifnum [1]{%
 \ifnum #1\expandafter \@firstoftwo
 \else \expandafter \@secondoftwo
 \fi
}%
\providecommand \@ifx [1]{%
 \ifx #1\expandafter \@firstoftwo
 \else \expandafter \@secondoftwo
 \fi
}%
\providecommand \natexlab [1]{#1}%
\providecommand \enquote  [1]{``#1''}%
\providecommand \bibnamefont  [1]{#1}%
\providecommand \bibfnamefont [1]{#1}%
\providecommand \citenamefont [1]{#1}%
\providecommand \href@noop [0]{\@secondoftwo}%
\providecommand \href [0]{\begingroup \@sanitize@url \@href}%
\providecommand \@href[1]{\@@startlink{#1}\@@href}%
\providecommand \@@href[1]{\endgroup#1\@@endlink}%
\providecommand \@sanitize@url [0]{\catcode `\\12\catcode `\$12\catcode
  `\&12\catcode `\#12\catcode `\^12\catcode `\_12\catcode `\%12\relax}%
\providecommand \@@startlink[1]{}%
\providecommand \@@endlink[0]{}%
\providecommand \url  [0]{\begingroup\@sanitize@url \@url }%
\providecommand \@url [1]{\endgroup\@href {#1}{\urlprefix }}%
\providecommand \urlprefix  [0]{URL }%
\providecommand \Eprint [0]{\href }%
\providecommand \doibase [0]{http://dx.doi.org/}%
\providecommand \selectlanguage [0]{\@gobble}%
\providecommand \bibinfo  [0]{\@secondoftwo}%
\providecommand \bibfield  [0]{\@secondoftwo}%
\providecommand \translation [1]{[#1]}%
\providecommand \BibitemOpen [0]{}%
\providecommand \bibitemStop [0]{}%
\providecommand \bibitemNoStop [0]{.\EOS\space}%
\providecommand \EOS [0]{\spacefactor3000\relax}%
\providecommand \BibitemShut  [1]{\csname bibitem#1\endcsname}%
\let\auto@bib@innerbib\@empty
\bibitem [{\citenamefont {Udem}\ \emph {et~al.}(2002)\citenamefont {Udem},
  \citenamefont {Holzwarth},\ and\ \citenamefont {Hänsch}}]{udem02}%
  \BibitemOpen
  \bibfield  {author} {\bibinfo {author} {\bibfnamefont {T.}~\bibnamefont
  {Udem}}, \bibinfo {author} {\bibfnamefont {R.}~\bibnamefont {Holzwarth}}, \
  and\ \bibinfo {author} {\bibfnamefont {T.~W.}\ \bibnamefont {Hänsch}},\
  }\href {\doibase 10.1038/416233a} {\bibfield  {journal} {\bibinfo  {journal}
  {Nature}\ }\textbf {\bibinfo {volume} {416}},\ \bibinfo {pages} {233}
  (\bibinfo {year} {2002})}\BibitemShut {NoStop}%
\bibitem [{\citenamefont {Cundiff}\ and\ \citenamefont {Ye}(2003)}]{Cundiff03}%
  \BibitemOpen
  \bibfield  {author} {\bibinfo {author} {\bibfnamefont {S.~T.}\ \bibnamefont
  {Cundiff}}\ and\ \bibinfo {author} {\bibfnamefont {J.}~\bibnamefont {Ye}},\
  }\href {https://link.aps.org/doi/10.1103/RevModPhys.75.325} {\bibfield
  {journal} {\bibinfo  {journal} {Rev. Mod. Phys.}\ }\textbf {\bibinfo {volume}
  {75}},\ \bibinfo {pages} {325} (\bibinfo {year} {2003})}\BibitemShut
  {NoStop}%
\bibitem [{\citenamefont {Diddams}(2010)}]{Diddams10}%
  \BibitemOpen
  \bibfield  {author} {\bibinfo {author} {\bibfnamefont {S.~A.}\ \bibnamefont
  {Diddams}},\ }\href {\doibase 10.1364/JOSAB.27.000B51} {\bibfield  {journal}
  {\bibinfo  {journal} {J. Opt. Soc. Am. B}\ }\textbf {\bibinfo {volume}
  {27}},\ \bibinfo {pages} {B51} (\bibinfo {year} {2010})}\BibitemShut
  {NoStop}%
\bibitem [{\citenamefont {Gaeta}\ \emph {et~al.}(2019)\citenamefont {Gaeta},
  \citenamefont {Lipson},\ and\ \citenamefont {Kippenberg}}]{Gaeta19}%
  \BibitemOpen
  \bibfield  {author} {\bibinfo {author} {\bibfnamefont {A.~L.}\ \bibnamefont
  {Gaeta}}, \bibinfo {author} {\bibfnamefont {M.}~\bibnamefont {Lipson}}, \
  and\ \bibinfo {author} {\bibfnamefont {T.~J.}\ \bibnamefont {Kippenberg}},\
  }\href {\doibase 10.1038/s41566-019-0358-x} {\bibfield  {journal} {\bibinfo
  {journal} {Nature Photonics}\ }\textbf {\bibinfo {volume} {13}},\ \bibinfo
  {pages} {158} (\bibinfo {year} {2019})}\BibitemShut {NoStop}%
\bibitem [{\citenamefont {Duchesne}\ \emph {et~al.}(2010)\citenamefont
  {Duchesne}, \citenamefont {Peccianti}, \citenamefont {Lamont}, \citenamefont
  {Ferrera}, \citenamefont {Razzari}, \citenamefont {L\'{e}gar\'{e}},
  \citenamefont {Morandotti}, \citenamefont {Chu}, \citenamefont {Little},\
  and\ \citenamefont {Moss}}]{Duchesne10}%
  \BibitemOpen
  \bibfield  {author} {\bibinfo {author} {\bibfnamefont {D.}~\bibnamefont
  {Duchesne}}, \bibinfo {author} {\bibfnamefont {M.}~\bibnamefont {Peccianti}},
  \bibinfo {author} {\bibfnamefont {M.~R.~E.}\ \bibnamefont {Lamont}}, \bibinfo
  {author} {\bibfnamefont {M.}~\bibnamefont {Ferrera}}, \bibinfo {author}
  {\bibfnamefont {L.}~\bibnamefont {Razzari}}, \bibinfo {author} {\bibfnamefont
  {F.}~\bibnamefont {L\'{e}gar\'{e}}}, \bibinfo {author} {\bibfnamefont
  {R.}~\bibnamefont {Morandotti}}, \bibinfo {author} {\bibfnamefont
  {S.}~\bibnamefont {Chu}}, \bibinfo {author} {\bibfnamefont {B.~E.}\
  \bibnamefont {Little}}, \ and\ \bibinfo {author} {\bibfnamefont {D.~J.}\
  \bibnamefont {Moss}},\ }\href {\doibase 10.1364/OE.18.000923} {\bibfield
  {journal} {\bibinfo  {journal} {Opt. Express}\ }\textbf {\bibinfo {volume}
  {18}},\ \bibinfo {pages} {923} (\bibinfo {year} {2010})}\BibitemShut
  {NoStop}%
\bibitem [{\citenamefont {Kuyken}\ \emph {et~al.}(2011)\citenamefont {Kuyken},
  \citenamefont {Liu}, \citenamefont {Osgood}, \citenamefont {Baets},
  \citenamefont {Roelkens},\ and\ \citenamefont {Green}}]{Kuyken11}%
  \BibitemOpen
  \bibfield  {author} {\bibinfo {author} {\bibfnamefont {B.}~\bibnamefont
  {Kuyken}}, \bibinfo {author} {\bibfnamefont {X.}~\bibnamefont {Liu}},
  \bibinfo {author} {\bibfnamefont {R.~M.}\ \bibnamefont {Osgood}}, \bibinfo
  {author} {\bibfnamefont {R.}~\bibnamefont {Baets}}, \bibinfo {author}
  {\bibfnamefont {G.}~\bibnamefont {Roelkens}}, \ and\ \bibinfo {author}
  {\bibfnamefont {W.~M.~J.}\ \bibnamefont {Green}},\ }\href {\doibase
  10.1364/OE.19.020172} {\bibfield  {journal} {\bibinfo  {journal} {Opt.
  Express}\ }\textbf {\bibinfo {volume} {19}},\ \bibinfo {pages} {20172}
  (\bibinfo {year} {2011})}\BibitemShut {NoStop}%
\bibitem [{\citenamefont {Halir}\ \emph {et~al.}(2012)\citenamefont {Halir},
  \citenamefont {Okawachi}, \citenamefont {Levy}, \citenamefont {Foster},
  \citenamefont {Lipson},\ and\ \citenamefont {Gaeta}}]{Halir12}%
  \BibitemOpen
  \bibfield  {author} {\bibinfo {author} {\bibfnamefont {R.}~\bibnamefont
  {Halir}}, \bibinfo {author} {\bibfnamefont {Y.}~\bibnamefont {Okawachi}},
  \bibinfo {author} {\bibfnamefont {J.~S.}\ \bibnamefont {Levy}}, \bibinfo
  {author} {\bibfnamefont {M.~A.}\ \bibnamefont {Foster}}, \bibinfo {author}
  {\bibfnamefont {M.}~\bibnamefont {Lipson}}, \ and\ \bibinfo {author}
  {\bibfnamefont {A.~L.}\ \bibnamefont {Gaeta}},\ }\href {\doibase
  10.1364/OL.37.001685} {\bibfield  {journal} {\bibinfo  {journal} {Opt.
  Lett.}\ }\textbf {\bibinfo {volume} {37}},\ \bibinfo {pages} {1685} (\bibinfo
  {year} {2012})}\BibitemShut {NoStop}%
\bibitem [{\citenamefont {Kuyken}\ \emph {et~al.}(2015)\citenamefont {Kuyken},
  \citenamefont {Ideguchi}, \citenamefont {Holzner}, \citenamefont {Yan},
  \citenamefont {Hänsch}, \citenamefont {Van~Campenhout}, \citenamefont
  {Verheyen}, \citenamefont {Coen}, \citenamefont {Leo}, \citenamefont {Baets},
  \citenamefont {Roelkens},\ and\ \citenamefont {Picqué}}]{Kuyken2015}%
  \BibitemOpen
  \bibfield  {author} {\bibinfo {author} {\bibfnamefont {B.}~\bibnamefont
  {Kuyken}}, \bibinfo {author} {\bibfnamefont {T.}~\bibnamefont {Ideguchi}},
  \bibinfo {author} {\bibfnamefont {S.}~\bibnamefont {Holzner}}, \bibinfo
  {author} {\bibfnamefont {M.}~\bibnamefont {Yan}}, \bibinfo {author}
  {\bibfnamefont {T.~W.}\ \bibnamefont {Hänsch}}, \bibinfo {author}
  {\bibfnamefont {J.}~\bibnamefont {Van~Campenhout}}, \bibinfo {author}
  {\bibfnamefont {P.}~\bibnamefont {Verheyen}}, \bibinfo {author}
  {\bibfnamefont {S.}~\bibnamefont {Coen}}, \bibinfo {author} {\bibfnamefont
  {F.}~\bibnamefont {Leo}}, \bibinfo {author} {\bibfnamefont {R.}~\bibnamefont
  {Baets}}, \bibinfo {author} {\bibfnamefont {G.}~\bibnamefont {Roelkens}}, \
  and\ \bibinfo {author} {\bibfnamefont {N.}~\bibnamefont {Picqué}},\ }\href
  {\doibase 10.1038/ncomms7310} {\bibfield  {journal} {\bibinfo  {journal}
  {Nature Communications}\ }\textbf {\bibinfo {volume} {6}},\ \bibinfo {pages}
  {6310} (\bibinfo {year} {2015})}\BibitemShut {NoStop}%
\bibitem [{\citenamefont {Singh}\ \emph {et~al.}(2015)\citenamefont {Singh},
  \citenamefont {Hudson}, \citenamefont {Yu}, \citenamefont {Grillet},
  \citenamefont {Jackson}, \citenamefont {Casas-Bedoya}, \citenamefont {Read},
  \citenamefont {Atanackovic}, \citenamefont {Duvall}, \citenamefont {Palomba},
  \citenamefont {Luther-Davies}, \citenamefont {Madden}, \citenamefont {Moss},\
  and\ \citenamefont {Eggleton}}]{Singh15}%
  \BibitemOpen
  \bibfield  {author} {\bibinfo {author} {\bibfnamefont {N.}~\bibnamefont
  {Singh}}, \bibinfo {author} {\bibfnamefont {D.~D.}\ \bibnamefont {Hudson}},
  \bibinfo {author} {\bibfnamefont {Y.}~\bibnamefont {Yu}}, \bibinfo {author}
  {\bibfnamefont {C.}~\bibnamefont {Grillet}}, \bibinfo {author} {\bibfnamefont
  {S.~D.}\ \bibnamefont {Jackson}}, \bibinfo {author} {\bibfnamefont
  {A.}~\bibnamefont {Casas-Bedoya}}, \bibinfo {author} {\bibfnamefont
  {A.}~\bibnamefont {Read}}, \bibinfo {author} {\bibfnamefont {P.}~\bibnamefont
  {Atanackovic}}, \bibinfo {author} {\bibfnamefont {S.~G.}\ \bibnamefont
  {Duvall}}, \bibinfo {author} {\bibfnamefont {S.}~\bibnamefont {Palomba}},
  \bibinfo {author} {\bibfnamefont {B.}~\bibnamefont {Luther-Davies}}, \bibinfo
  {author} {\bibfnamefont {S.}~\bibnamefont {Madden}}, \bibinfo {author}
  {\bibfnamefont {D.~J.}\ \bibnamefont {Moss}}, \ and\ \bibinfo {author}
  {\bibfnamefont {B.~J.}\ \bibnamefont {Eggleton}},\ }\href {\doibase
  10.1364/OPTICA.2.000797} {\bibfield  {journal} {\bibinfo  {journal} {Optica}\
  }\textbf {\bibinfo {volume} {2}},\ \bibinfo {pages} {797} (\bibinfo {year}
  {2015})}\BibitemShut {NoStop}%
\bibitem [{\citenamefont {Liu}\ \emph {et~al.}(2016)\citenamefont {Liu},
  \citenamefont {Pu}, \citenamefont {Zhou}, \citenamefont {Kr\"{u}ckel},
  \citenamefont {F\"{u}l\"{o}p}, \citenamefont {Torres-Company},\ and\
  \citenamefont {Bache}}]{Liu16}%
  \BibitemOpen
  \bibfield  {author} {\bibinfo {author} {\bibfnamefont {X.}~\bibnamefont
  {Liu}}, \bibinfo {author} {\bibfnamefont {M.}~\bibnamefont {Pu}}, \bibinfo
  {author} {\bibfnamefont {B.}~\bibnamefont {Zhou}}, \bibinfo {author}
  {\bibfnamefont {C.~J.}\ \bibnamefont {Kr\"{u}ckel}}, \bibinfo {author}
  {\bibfnamefont {A.}~\bibnamefont {F\"{u}l\"{o}p}}, \bibinfo {author}
  {\bibfnamefont {V.}~\bibnamefont {Torres-Company}}, \ and\ \bibinfo {author}
  {\bibfnamefont {M.}~\bibnamefont {Bache}},\ }\href {\doibase
  10.1364/OL.41.002719} {\bibfield  {journal} {\bibinfo  {journal} {Opt.
  Lett.}\ }\textbf {\bibinfo {volume} {41}},\ \bibinfo {pages} {2719} (\bibinfo
  {year} {2016})}\BibitemShut {NoStop}%
\bibitem [{\citenamefont {Klenner}\ \emph {et~al.}(2016)\citenamefont
  {Klenner}, \citenamefont {Mayer}, \citenamefont {Johnson}, \citenamefont
  {Luke}, \citenamefont {Lamont}, \citenamefont {Okawachi}, \citenamefont
  {Lipson}, \citenamefont {Gaeta},\ and\ \citenamefont {Keller}}]{Klenner16}%
  \BibitemOpen
  \bibfield  {author} {\bibinfo {author} {\bibfnamefont {A.}~\bibnamefont
  {Klenner}}, \bibinfo {author} {\bibfnamefont {A.~S.}\ \bibnamefont {Mayer}},
  \bibinfo {author} {\bibfnamefont {A.~R.}\ \bibnamefont {Johnson}}, \bibinfo
  {author} {\bibfnamefont {K.}~\bibnamefont {Luke}}, \bibinfo {author}
  {\bibfnamefont {M.~R.~E.}\ \bibnamefont {Lamont}}, \bibinfo {author}
  {\bibfnamefont {Y.}~\bibnamefont {Okawachi}}, \bibinfo {author}
  {\bibfnamefont {M.}~\bibnamefont {Lipson}}, \bibinfo {author} {\bibfnamefont
  {A.~L.}\ \bibnamefont {Gaeta}}, \ and\ \bibinfo {author} {\bibfnamefont
  {U.}~\bibnamefont {Keller}},\ }\href {\doibase 10.1364/OE.24.011043}
  {\bibfield  {journal} {\bibinfo  {journal} {Opt. Express}\ }\textbf {\bibinfo
  {volume} {24}},\ \bibinfo {pages} {11043} (\bibinfo {year}
  {2016})}\BibitemShut {NoStop}%
\bibitem [{\citenamefont {Carlson}\ \emph {et~al.}(2017)\citenamefont
  {Carlson}, \citenamefont {Hickstein}, \citenamefont {Lind}, \citenamefont
  {Droste}, \citenamefont {Westly}, \citenamefont {Nader}, \citenamefont
  {Coddington}, \citenamefont {Newbury}, \citenamefont {Srinivasan},
  \citenamefont {Diddams},\ and\ \citenamefont {Papp}}]{Carlson17}%
  \BibitemOpen
  \bibfield  {author} {\bibinfo {author} {\bibfnamefont {D.~R.}\ \bibnamefont
  {Carlson}}, \bibinfo {author} {\bibfnamefont {D.~D.}\ \bibnamefont
  {Hickstein}}, \bibinfo {author} {\bibfnamefont {A.}~\bibnamefont {Lind}},
  \bibinfo {author} {\bibfnamefont {S.}~\bibnamefont {Droste}}, \bibinfo
  {author} {\bibfnamefont {D.}~\bibnamefont {Westly}}, \bibinfo {author}
  {\bibfnamefont {N.}~\bibnamefont {Nader}}, \bibinfo {author} {\bibfnamefont
  {I.}~\bibnamefont {Coddington}}, \bibinfo {author} {\bibfnamefont {N.~R.}\
  \bibnamefont {Newbury}}, \bibinfo {author} {\bibfnamefont {K.}~\bibnamefont
  {Srinivasan}}, \bibinfo {author} {\bibfnamefont {S.~A.}\ \bibnamefont
  {Diddams}}, \ and\ \bibinfo {author} {\bibfnamefont {S.~B.}\ \bibnamefont
  {Papp}},\ }\href {\doibase 10.1364/OL.42.002314} {\bibfield  {journal}
  {\bibinfo  {journal} {Opt. Lett.}\ }\textbf {\bibinfo {volume} {42}},\
  \bibinfo {pages} {2314} (\bibinfo {year} {2017})}\BibitemShut {NoStop}%
\bibitem [{\citenamefont {Porcel}\ \emph {et~al.}(2017)\citenamefont {Porcel},
  \citenamefont {Schepers}, \citenamefont {Epping}, \citenamefont {Hellwig},
  \citenamefont {Hoekman}, \citenamefont {Heideman}, \citenamefont {van~der
  Slot}, \citenamefont {Lee}, \citenamefont {Schmidt}, \citenamefont
  {Bratschitsch}, \citenamefont {Fallnich},\ and\ \citenamefont
  {Boller}}]{Porcel17}%
  \BibitemOpen
  \bibfield  {author} {\bibinfo {author} {\bibfnamefont {M.~A.~G.}\
  \bibnamefont {Porcel}}, \bibinfo {author} {\bibfnamefont {F.}~\bibnamefont
  {Schepers}}, \bibinfo {author} {\bibfnamefont {J.~P.}\ \bibnamefont
  {Epping}}, \bibinfo {author} {\bibfnamefont {T.}~\bibnamefont {Hellwig}},
  \bibinfo {author} {\bibfnamefont {M.}~\bibnamefont {Hoekman}}, \bibinfo
  {author} {\bibfnamefont {R.~G.}\ \bibnamefont {Heideman}}, \bibinfo {author}
  {\bibfnamefont {P.~J.~M.}\ \bibnamefont {van~der Slot}}, \bibinfo {author}
  {\bibfnamefont {C.~J.}\ \bibnamefont {Lee}}, \bibinfo {author} {\bibfnamefont
  {R.}~\bibnamefont {Schmidt}}, \bibinfo {author} {\bibfnamefont
  {R.}~\bibnamefont {Bratschitsch}}, \bibinfo {author} {\bibfnamefont
  {C.}~\bibnamefont {Fallnich}}, \ and\ \bibinfo {author} {\bibfnamefont
  {K.-J.}\ \bibnamefont {Boller}},\ }\href {\doibase 10.1364/OE.25.001542}
  {\bibfield  {journal} {\bibinfo  {journal} {Opt. Express}\ }\textbf {\bibinfo
  {volume} {25}},\ \bibinfo {pages} {1542} (\bibinfo {year}
  {2017})}\BibitemShut {NoStop}%
\bibitem [{\citenamefont {Okawachi}\ \emph {et~al.}(2018)\citenamefont
  {Okawachi}, \citenamefont {Yu}, \citenamefont {Cardenas}, \citenamefont {Ji},
  \citenamefont {Klenner}, \citenamefont {Lipson},\ and\ \citenamefont
  {Gaeta}}]{Okawachi18}%
  \BibitemOpen
  \bibfield  {author} {\bibinfo {author} {\bibfnamefont {Y.}~\bibnamefont
  {Okawachi}}, \bibinfo {author} {\bibfnamefont {M.}~\bibnamefont {Yu}},
  \bibinfo {author} {\bibfnamefont {J.}~\bibnamefont {Cardenas}}, \bibinfo
  {author} {\bibfnamefont {X.}~\bibnamefont {Ji}}, \bibinfo {author}
  {\bibfnamefont {A.}~\bibnamefont {Klenner}}, \bibinfo {author} {\bibfnamefont
  {M.}~\bibnamefont {Lipson}}, \ and\ \bibinfo {author} {\bibfnamefont {A.~L.}\
  \bibnamefont {Gaeta}},\ }\href {\doibase 10.1364/OL.43.004627} {\bibfield
  {journal} {\bibinfo  {journal} {Opt. Lett.}\ }\textbf {\bibinfo {volume}
  {43}},\ \bibinfo {pages} {4627} (\bibinfo {year} {2018})}\BibitemShut
  {NoStop}%
\bibitem [{\citenamefont {Waldburger}\ \emph {et~al.}(2019)\citenamefont
  {Waldburger}, \citenamefont {Mayer}, \citenamefont {Alfieri}, \citenamefont
  {N\"{u}rnberg}, \citenamefont {Johnson}, \citenamefont {Ji}, \citenamefont
  {Klenner}, \citenamefont {Okawachi}, \citenamefont {Lipson}, \citenamefont
  {Gaeta},\ and\ \citenamefont {Keller}}]{Waldburger19}%
  \BibitemOpen
  \bibfield  {author} {\bibinfo {author} {\bibfnamefont {D.}~\bibnamefont
  {Waldburger}}, \bibinfo {author} {\bibfnamefont {A.~S.}\ \bibnamefont
  {Mayer}}, \bibinfo {author} {\bibfnamefont {C.~G.~E.}\ \bibnamefont
  {Alfieri}}, \bibinfo {author} {\bibfnamefont {J.}~\bibnamefont
  {N\"{u}rnberg}}, \bibinfo {author} {\bibfnamefont {A.~R.}\ \bibnamefont
  {Johnson}}, \bibinfo {author} {\bibfnamefont {X.}~\bibnamefont {Ji}},
  \bibinfo {author} {\bibfnamefont {A.}~\bibnamefont {Klenner}}, \bibinfo
  {author} {\bibfnamefont {Y.}~\bibnamefont {Okawachi}}, \bibinfo {author}
  {\bibfnamefont {M.}~\bibnamefont {Lipson}}, \bibinfo {author} {\bibfnamefont
  {A.~L.}\ \bibnamefont {Gaeta}}, \ and\ \bibinfo {author} {\bibfnamefont
  {U.}~\bibnamefont {Keller}},\ }\href {\doibase 10.1364/OE.27.001786}
  {\bibfield  {journal} {\bibinfo  {journal} {Opt. Express}\ }\textbf {\bibinfo
  {volume} {27}},\ \bibinfo {pages} {1786} (\bibinfo {year}
  {2019})}\BibitemShut {NoStop}%
\bibitem [{\citenamefont {Hickstein}\ \emph {et~al.}(2019)\citenamefont
  {Hickstein}, \citenamefont {Carlson}, \citenamefont {Mundoor}, \citenamefont
  {Khurgin}, \citenamefont {Srinivasan}, \citenamefont {Westly}, \citenamefont
  {Kowligy}, \citenamefont {Smalyukh}, \citenamefont {Diddams},\ and\
  \citenamefont {Papp}}]{Hickstein19}%
  \BibitemOpen
  \bibfield  {author} {\bibinfo {author} {\bibfnamefont {D.~D.}\ \bibnamefont
  {Hickstein}}, \bibinfo {author} {\bibfnamefont {D.~R.}\ \bibnamefont
  {Carlson}}, \bibinfo {author} {\bibfnamefont {H.}~\bibnamefont {Mundoor}},
  \bibinfo {author} {\bibfnamefont {J.~B.}\ \bibnamefont {Khurgin}}, \bibinfo
  {author} {\bibfnamefont {K.}~\bibnamefont {Srinivasan}}, \bibinfo {author}
  {\bibfnamefont {D.}~\bibnamefont {Westly}}, \bibinfo {author} {\bibfnamefont
  {A.}~\bibnamefont {Kowligy}}, \bibinfo {author} {\bibfnamefont {I.~I.}\
  \bibnamefont {Smalyukh}}, \bibinfo {author} {\bibfnamefont {S.~A.}\
  \bibnamefont {Diddams}}, \ and\ \bibinfo {author} {\bibfnamefont {S.~B.}\
  \bibnamefont {Papp}},\ }\href {\doibase 10.1038/s41566-019-0449-8} {\bibfield
   {journal} {\bibinfo  {journal} {Nature Photonics}\ }\textbf {\bibinfo
  {volume} {13}},\ \bibinfo {pages} {494} (\bibinfo {year} {2019})}\BibitemShut
  {NoStop}%
\bibitem [{\citenamefont {Yoon~Oh}\ \emph {et~al.}(2017)\citenamefont
  {Yoon~Oh}, \citenamefont {Yang}, \citenamefont {Fredrick}, \citenamefont
  {Ycas}, \citenamefont {Diddams},\ and\ \citenamefont {Vahala}}]{Oh17}%
  \BibitemOpen
  \bibfield  {author} {\bibinfo {author} {\bibfnamefont {D.}~\bibnamefont
  {Yoon~Oh}}, \bibinfo {author} {\bibfnamefont {K.~Y.}\ \bibnamefont {Yang}},
  \bibinfo {author} {\bibfnamefont {C.}~\bibnamefont {Fredrick}}, \bibinfo
  {author} {\bibfnamefont {G.}~\bibnamefont {Ycas}}, \bibinfo {author}
  {\bibfnamefont {S.~A.}\ \bibnamefont {Diddams}}, \ and\ \bibinfo {author}
  {\bibfnamefont {K.~J.}\ \bibnamefont {Vahala}},\ }\href {\doibase
  10.1038/ncomms13922} {\bibfield  {journal} {\bibinfo  {journal} {Nature
  Communications}\ }\textbf {\bibinfo {volume} {8}},\ \bibinfo {pages} {13922}
  (\bibinfo {year} {2017})}\BibitemShut {NoStop}%
\bibitem [{\citenamefont {Singh}\ \emph {et~al.}(2018)\citenamefont {Singh},
  \citenamefont {Xin}, \citenamefont {Vermeulen}, \citenamefont {Shtyrkova},
  \citenamefont {Li}, \citenamefont {Callahan}, \citenamefont {Magden},
  \citenamefont {Ruocco}, \citenamefont {Fahrenkopf}, \citenamefont {Baiocco},
  \citenamefont {Kuo}, \citenamefont {Radic}, \citenamefont {Ippen},
  \citenamefont {Kärtner},\ and\ \citenamefont {Watts}}]{Singh18}%
  \BibitemOpen
  \bibfield  {author} {\bibinfo {author} {\bibfnamefont {N.}~\bibnamefont
  {Singh}}, \bibinfo {author} {\bibfnamefont {M.}~\bibnamefont {Xin}}, \bibinfo
  {author} {\bibfnamefont {D.}~\bibnamefont {Vermeulen}}, \bibinfo {author}
  {\bibfnamefont {K.}~\bibnamefont {Shtyrkova}}, \bibinfo {author}
  {\bibfnamefont {N.}~\bibnamefont {Li}}, \bibinfo {author} {\bibfnamefont
  {P.~T.}\ \bibnamefont {Callahan}}, \bibinfo {author} {\bibfnamefont {E.~S.}\
  \bibnamefont {Magden}}, \bibinfo {author} {\bibfnamefont {A.}~\bibnamefont
  {Ruocco}}, \bibinfo {author} {\bibfnamefont {N.}~\bibnamefont {Fahrenkopf}},
  \bibinfo {author} {\bibfnamefont {C.}~\bibnamefont {Baiocco}}, \bibinfo
  {author} {\bibfnamefont {B.~P.-P.}\ \bibnamefont {Kuo}}, \bibinfo {author}
  {\bibfnamefont {S.}~\bibnamefont {Radic}}, \bibinfo {author} {\bibfnamefont
  {E.}~\bibnamefont {Ippen}}, \bibinfo {author} {\bibfnamefont {F.~X.}\
  \bibnamefont {Kärtner}}, \ and\ \bibinfo {author} {\bibfnamefont {M.~R.}\
  \bibnamefont {Watts}},\ }\href {\doibase 10.1038/lsa.2017.131} {\bibfield
  {journal} {\bibinfo  {journal} {Light: Science \& Applications}\ }\textbf
  {\bibinfo {volume} {7}},\ \bibinfo {pages} {17131} (\bibinfo {year}
  {2018})}\BibitemShut {NoStop}%
\bibitem [{\citenamefont {Hickstein}\ \emph {et~al.}(2017)\citenamefont
  {Hickstein}, \citenamefont {Jung}, \citenamefont {Carlson}, \citenamefont
  {Lind}, \citenamefont {Coddington}, \citenamefont {Srinivasan}, \citenamefont
  {Ycas}, \citenamefont {Cole}, \citenamefont {Kowligy}, \citenamefont
  {Fredrick}, \citenamefont {Droste}, \citenamefont {Lamb}, \citenamefont
  {Newbury}, \citenamefont {Tang}, \citenamefont {Diddams},\ and\ \citenamefont
  {Papp}}]{Hickstein17}%
  \BibitemOpen
  \bibfield  {author} {\bibinfo {author} {\bibfnamefont {D.~D.}\ \bibnamefont
  {Hickstein}}, \bibinfo {author} {\bibfnamefont {H.}~\bibnamefont {Jung}},
  \bibinfo {author} {\bibfnamefont {D.~R.}\ \bibnamefont {Carlson}}, \bibinfo
  {author} {\bibfnamefont {A.}~\bibnamefont {Lind}}, \bibinfo {author}
  {\bibfnamefont {I.}~\bibnamefont {Coddington}}, \bibinfo {author}
  {\bibfnamefont {K.}~\bibnamefont {Srinivasan}}, \bibinfo {author}
  {\bibfnamefont {G.~G.}\ \bibnamefont {Ycas}}, \bibinfo {author}
  {\bibfnamefont {D.~C.}\ \bibnamefont {Cole}}, \bibinfo {author}
  {\bibfnamefont {A.}~\bibnamefont {Kowligy}}, \bibinfo {author} {\bibfnamefont
  {C.}~\bibnamefont {Fredrick}}, \bibinfo {author} {\bibfnamefont
  {S.}~\bibnamefont {Droste}}, \bibinfo {author} {\bibfnamefont {E.~S.}\
  \bibnamefont {Lamb}}, \bibinfo {author} {\bibfnamefont {N.~R.}\ \bibnamefont
  {Newbury}}, \bibinfo {author} {\bibfnamefont {H.~X.}\ \bibnamefont {Tang}},
  \bibinfo {author} {\bibfnamefont {S.~A.}\ \bibnamefont {Diddams}}, \ and\
  \bibinfo {author} {\bibfnamefont {S.~B.}\ \bibnamefont {Papp}},\ }\href
  {\doibase 10.1103/PhysRevApplied.8.014025} {\bibfield  {journal} {\bibinfo
  {journal} {Phys. Rev. Applied}\ }\textbf {\bibinfo {volume} {8}},\ \bibinfo
  {pages} {014025} (\bibinfo {year} {2017})}\BibitemShut {NoStop}%
\bibitem [{\citenamefont {Pfeiffer}\ \emph {et~al.}(2017)\citenamefont
  {Pfeiffer}, \citenamefont {Herkommer}, \citenamefont {Liu}, \citenamefont
  {Guo}, \citenamefont {Karpov}, \citenamefont {Lucas}, \citenamefont
  {Zervas},\ and\ \citenamefont {Kippenberg}}]{Pfeiffer17}%
  \BibitemOpen
  \bibfield  {author} {\bibinfo {author} {\bibfnamefont {M.~H.~P.}\
  \bibnamefont {Pfeiffer}}, \bibinfo {author} {\bibfnamefont {C.}~\bibnamefont
  {Herkommer}}, \bibinfo {author} {\bibfnamefont {J.}~\bibnamefont {Liu}},
  \bibinfo {author} {\bibfnamefont {H.}~\bibnamefont {Guo}}, \bibinfo {author}
  {\bibfnamefont {M.}~\bibnamefont {Karpov}}, \bibinfo {author} {\bibfnamefont
  {E.}~\bibnamefont {Lucas}}, \bibinfo {author} {\bibfnamefont
  {M.}~\bibnamefont {Zervas}}, \ and\ \bibinfo {author} {\bibfnamefont {T.~J.}\
  \bibnamefont {Kippenberg}},\ }\href {\doibase 10.1364/OPTICA.4.000684}
  {\bibfield  {journal} {\bibinfo  {journal} {Optica}\ }\textbf {\bibinfo
  {volume} {4}},\ \bibinfo {pages} {684} (\bibinfo {year} {2017})}\BibitemShut
  {NoStop}%
\bibitem [{\citenamefont {Brasch}\ \emph {et~al.}(2017)\citenamefont {Brasch},
  \citenamefont {Lucas}, \citenamefont {Jost}, \citenamefont {Geiselmann},\
  and\ \citenamefont {Kippenberg}}]{Brasch17}%
  \BibitemOpen
  \bibfield  {author} {\bibinfo {author} {\bibfnamefont {V.}~\bibnamefont
  {Brasch}}, \bibinfo {author} {\bibfnamefont {E.}~\bibnamefont {Lucas}},
  \bibinfo {author} {\bibfnamefont {J.~D.}\ \bibnamefont {Jost}}, \bibinfo
  {author} {\bibfnamefont {M.}~\bibnamefont {Geiselmann}}, \ and\ \bibinfo
  {author} {\bibfnamefont {T.~J.}\ \bibnamefont {Kippenberg}},\ }\href
  {\doibase 10.1038/lsa.2016.202} {\bibfield  {journal} {\bibinfo  {journal}
  {Light: Science \& Applications}\ }\textbf {\bibinfo {volume} {6}},\ \bibinfo
  {pages} {e16202} (\bibinfo {year} {2017})}\BibitemShut {NoStop}%
\bibitem [{\citenamefont {Drake}\ \emph {et~al.}(2019)\citenamefont {Drake},
  \citenamefont {Briles}, \citenamefont {Stone}, \citenamefont {Spencer},
  \citenamefont {Carlson}, \citenamefont {Hickstein}, \citenamefont {Li},
  \citenamefont {Westly}, \citenamefont {Srinivasan}, \citenamefont {Diddams},\
  and\ \citenamefont {Papp}}]{Drake19}%
  \BibitemOpen
  \bibfield  {author} {\bibinfo {author} {\bibfnamefont {T.~E.}\ \bibnamefont
  {Drake}}, \bibinfo {author} {\bibfnamefont {T.~C.}\ \bibnamefont {Briles}},
  \bibinfo {author} {\bibfnamefont {J.~R.}\ \bibnamefont {Stone}}, \bibinfo
  {author} {\bibfnamefont {D.~T.}\ \bibnamefont {Spencer}}, \bibinfo {author}
  {\bibfnamefont {D.~R.}\ \bibnamefont {Carlson}}, \bibinfo {author}
  {\bibfnamefont {D.~D.}\ \bibnamefont {Hickstein}}, \bibinfo {author}
  {\bibfnamefont {Q.}~\bibnamefont {Li}}, \bibinfo {author} {\bibfnamefont
  {D.}~\bibnamefont {Westly}}, \bibinfo {author} {\bibfnamefont
  {K.}~\bibnamefont {Srinivasan}}, \bibinfo {author} {\bibfnamefont {S.~A.}\
  \bibnamefont {Diddams}}, \ and\ \bibinfo {author} {\bibfnamefont {S.~B.}\
  \bibnamefont {Papp}},\ }\href {\doibase 10.1103/PhysRevX.9.031023} {\bibfield
   {journal} {\bibinfo  {journal} {Phys. Rev. X}\ }\textbf {\bibinfo {volume}
  {9}},\ \bibinfo {pages} {031023} (\bibinfo {year} {2019})}\BibitemShut
  {NoStop}%
\bibitem [{\citenamefont {Newman}\ \emph {et~al.}(2019)\citenamefont {Newman},
  \citenamefont {Maurice}, \citenamefont {Drake}, \citenamefont {Stone},
  \citenamefont {Briles}, \citenamefont {Spencer}, \citenamefont {Fredrick},
  \citenamefont {Li}, \citenamefont {Westly}, \citenamefont {Ilic},
  \citenamefont {Shen}, \citenamefont {Suh}, \citenamefont {Yang},
  \citenamefont {Johnson}, \citenamefont {Johnson}, \citenamefont {Hollberg},
  \citenamefont {Vahala}, \citenamefont {Srinivasan}, \citenamefont {Diddams},
  \citenamefont {Kitching}, \citenamefont {Papp},\ and\ \citenamefont
  {Hummon}}]{Newman19}%
  \BibitemOpen
  \bibfield  {author} {\bibinfo {author} {\bibfnamefont {Z.~L.}\ \bibnamefont
  {Newman}}, \bibinfo {author} {\bibfnamefont {V.}~\bibnamefont {Maurice}},
  \bibinfo {author} {\bibfnamefont {T.}~\bibnamefont {Drake}}, \bibinfo
  {author} {\bibfnamefont {J.~R.}\ \bibnamefont {Stone}}, \bibinfo {author}
  {\bibfnamefont {T.~C.}\ \bibnamefont {Briles}}, \bibinfo {author}
  {\bibfnamefont {D.~T.}\ \bibnamefont {Spencer}}, \bibinfo {author}
  {\bibfnamefont {C.}~\bibnamefont {Fredrick}}, \bibinfo {author}
  {\bibfnamefont {Q.}~\bibnamefont {Li}}, \bibinfo {author} {\bibfnamefont
  {D.}~\bibnamefont {Westly}}, \bibinfo {author} {\bibfnamefont {B.~R.}\
  \bibnamefont {Ilic}}, \bibinfo {author} {\bibfnamefont {B.}~\bibnamefont
  {Shen}}, \bibinfo {author} {\bibfnamefont {M.-G.}\ \bibnamefont {Suh}},
  \bibinfo {author} {\bibfnamefont {K.~Y.}\ \bibnamefont {Yang}}, \bibinfo
  {author} {\bibfnamefont {C.}~\bibnamefont {Johnson}}, \bibinfo {author}
  {\bibfnamefont {D.~M.~S.}\ \bibnamefont {Johnson}}, \bibinfo {author}
  {\bibfnamefont {L.}~\bibnamefont {Hollberg}}, \bibinfo {author}
  {\bibfnamefont {K.~J.}\ \bibnamefont {Vahala}}, \bibinfo {author}
  {\bibfnamefont {K.}~\bibnamefont {Srinivasan}}, \bibinfo {author}
  {\bibfnamefont {S.~A.}\ \bibnamefont {Diddams}}, \bibinfo {author}
  {\bibfnamefont {J.}~\bibnamefont {Kitching}}, \bibinfo {author}
  {\bibfnamefont {S.~B.}\ \bibnamefont {Papp}}, \ and\ \bibinfo {author}
  {\bibfnamefont {M.~T.}\ \bibnamefont {Hummon}},\ }\href {\doibase
  10.1364/OPTICA.6.000680} {\bibfield  {journal} {\bibinfo  {journal} {Optica}\
  }\textbf {\bibinfo {volume} {6}},\ \bibinfo {pages} {680} (\bibinfo {year}
  {2019})}\BibitemShut {NoStop}%
\bibitem [{\citenamefont {Telle}\ \emph {et~al.}(1999)\citenamefont {Telle},
  \citenamefont {Steinmeyer}, \citenamefont {Dunlop}, \citenamefont {Stenger},
  \citenamefont {Sutter},\ and\ \citenamefont {Keller}}]{Telle99}%
  \BibitemOpen
  \bibfield  {author} {\bibinfo {author} {\bibfnamefont {H.}~\bibnamefont
  {Telle}}, \bibinfo {author} {\bibfnamefont {G.}~\bibnamefont {Steinmeyer}},
  \bibinfo {author} {\bibfnamefont {A.}~\bibnamefont {Dunlop}}, \bibinfo
  {author} {\bibfnamefont {J.}~\bibnamefont {Stenger}}, \bibinfo {author}
  {\bibfnamefont {D.}~\bibnamefont {Sutter}}, \ and\ \bibinfo {author}
  {\bibfnamefont {U.}~\bibnamefont {Keller}},\ }\href {\doibase
  10.1007/s003400050813} {\bibfield  {journal} {\bibinfo  {journal} {Appl.
  Phys. B}\ }\textbf {\bibinfo {volume} {69}},\ \bibinfo {pages} {327}
  (\bibinfo {year} {1999})}\BibitemShut {NoStop}%
\bibitem [{\citenamefont {Jones}\ \emph {et~al.}(2000)\citenamefont {Jones},
  \citenamefont {Diddams}, \citenamefont {Ranka}, \citenamefont {Stentz},
  \citenamefont {Windeler}, \citenamefont {Hall},\ and\ \citenamefont
  {Cundiff}}]{Jones00}%
  \BibitemOpen
  \bibfield  {author} {\bibinfo {author} {\bibfnamefont {D.~J.}\ \bibnamefont
  {Jones}}, \bibinfo {author} {\bibfnamefont {S.~A.}\ \bibnamefont {Diddams}},
  \bibinfo {author} {\bibfnamefont {J.~K.}\ \bibnamefont {Ranka}}, \bibinfo
  {author} {\bibfnamefont {A.}~\bibnamefont {Stentz}}, \bibinfo {author}
  {\bibfnamefont {R.~S.}\ \bibnamefont {Windeler}}, \bibinfo {author}
  {\bibfnamefont {J.~L.}\ \bibnamefont {Hall}}, \ and\ \bibinfo {author}
  {\bibfnamefont {S.~T.}\ \bibnamefont {Cundiff}},\ }\href {\doibase
  10.1126/science.288.5466.635} {\bibfield  {journal} {\bibinfo  {journal}
  {Science}\ }\textbf {\bibinfo {volume} {288}},\ \bibinfo {pages} {635}
  (\bibinfo {year} {2000})},\ \Eprint
  {http://arxiv.org/abs/https://science.sciencemag.org/content/288/5466/635.full.pdf}
  {https://science.sciencemag.org/content/288/5466/635.full.pdf} \BibitemShut
  {NoStop}%
\bibitem [{\citenamefont {Holzwarth}\ \emph {et~al.}(2000)\citenamefont
  {Holzwarth}, \citenamefont {Udem}, \citenamefont {H\"ansch}, \citenamefont
  {Knight}, \citenamefont {Wadsworth},\ and\ \citenamefont
  {Russell}}]{Holzwarth00}%
  \BibitemOpen
  \bibfield  {author} {\bibinfo {author} {\bibfnamefont {R.}~\bibnamefont
  {Holzwarth}}, \bibinfo {author} {\bibfnamefont {T.}~\bibnamefont {Udem}},
  \bibinfo {author} {\bibfnamefont {T.~W.}\ \bibnamefont {H\"ansch}}, \bibinfo
  {author} {\bibfnamefont {J.~C.}\ \bibnamefont {Knight}}, \bibinfo {author}
  {\bibfnamefont {W.~J.}\ \bibnamefont {Wadsworth}}, \ and\ \bibinfo {author}
  {\bibfnamefont {P.~S.~J.}\ \bibnamefont {Russell}},\ }\href {\doibase
  10.1103/PhysRevLett.85.2264} {\bibfield  {journal} {\bibinfo  {journal}
  {Phys. Rev. Lett.}\ }\textbf {\bibinfo {volume} {85}},\ \bibinfo {pages}
  {2264} (\bibinfo {year} {2000})}\BibitemShut {NoStop}%
\bibitem [{\citenamefont {Wang}\ \emph {et~al.}(2019)\citenamefont {Wang},
  \citenamefont {Zhang}, \citenamefont {Yu}, \citenamefont {Zhu}, \citenamefont
  {Hu},\ and\ \citenamefont {Loncar}}]{WangComm}%
  \BibitemOpen
  \bibfield  {author} {\bibinfo {author} {\bibfnamefont {C.}~\bibnamefont
  {Wang}}, \bibinfo {author} {\bibfnamefont {M.}~\bibnamefont {Zhang}},
  \bibinfo {author} {\bibfnamefont {M.}~\bibnamefont {Yu}}, \bibinfo {author}
  {\bibfnamefont {R.}~\bibnamefont {Zhu}}, \bibinfo {author} {\bibfnamefont
  {H.}~\bibnamefont {Hu}}, \ and\ \bibinfo {author} {\bibfnamefont
  {M.}~\bibnamefont {Loncar}},\ }\href {\doibase 10.1038/s41467-019-08969-6}
  {\bibfield  {journal} {\bibinfo  {journal} {Nature Communications}\ }\textbf
  {\bibinfo {volume} {10}},\ \bibinfo {pages} {978} (\bibinfo {year}
  {2019})}\BibitemShut {NoStop}%
\bibitem [{\citenamefont {Wang}\ \emph {et~al.}(2018)\citenamefont {Wang},
  \citenamefont {Langrock}, \citenamefont {Marandi}, \citenamefont {Jankowski},
  \citenamefont {Zhang}, \citenamefont {Desiatov}, \citenamefont {Fejer},\ and\
  \citenamefont {Lon\v{c}ar}}]{WangOptica}%
  \BibitemOpen
  \bibfield  {author} {\bibinfo {author} {\bibfnamefont {C.}~\bibnamefont
  {Wang}}, \bibinfo {author} {\bibfnamefont {C.}~\bibnamefont {Langrock}},
  \bibinfo {author} {\bibfnamefont {A.}~\bibnamefont {Marandi}}, \bibinfo
  {author} {\bibfnamefont {M.}~\bibnamefont {Jankowski}}, \bibinfo {author}
  {\bibfnamefont {M.}~\bibnamefont {Zhang}}, \bibinfo {author} {\bibfnamefont
  {B.}~\bibnamefont {Desiatov}}, \bibinfo {author} {\bibfnamefont {M.~M.}\
  \bibnamefont {Fejer}}, \ and\ \bibinfo {author} {\bibfnamefont
  {M.}~\bibnamefont {Lon\v{c}ar}},\ }\href {\doibase 10.1364/OPTICA.5.001438}
  {\bibfield  {journal} {\bibinfo  {journal} {Optica}\ }\textbf {\bibinfo
  {volume} {5}},\ \bibinfo {pages} {1438} (\bibinfo {year} {2018})}\BibitemShut
  {NoStop}%
\bibitem [{\citenamefont {Yu}\ \emph {et~al.}(2019)\citenamefont {Yu},
  \citenamefont {Desiatov}, \citenamefont {Okawachi}, \citenamefont {Gaeta},\
  and\ \citenamefont {Lon\v{c}ar}}]{Yu19}%
  \BibitemOpen
  \bibfield  {author} {\bibinfo {author} {\bibfnamefont {M.}~\bibnamefont
  {Yu}}, \bibinfo {author} {\bibfnamefont {B.}~\bibnamefont {Desiatov}},
  \bibinfo {author} {\bibfnamefont {Y.}~\bibnamefont {Okawachi}}, \bibinfo
  {author} {\bibfnamefont {A.~L.}\ \bibnamefont {Gaeta}}, \ and\ \bibinfo
  {author} {\bibfnamefont {M.}~\bibnamefont {Lon\v{c}ar}},\ }\href {\doibase
  10.1364/OL.44.001222} {\bibfield  {journal} {\bibinfo  {journal} {Opt.
  Lett.}\ }\textbf {\bibinfo {volume} {44}},\ \bibinfo {pages} {1222} (\bibinfo
  {year} {2019})}\BibitemShut {NoStop}%
\bibitem [{\citenamefont {Lu}\ \emph {et~al.}(2019)\citenamefont {Lu},
  \citenamefont {Surya}, \citenamefont {Liu}, \citenamefont {Xu},\ and\
  \citenamefont {Tang}}]{Lu}%
  \BibitemOpen
  \bibfield  {author} {\bibinfo {author} {\bibfnamefont {J.}~\bibnamefont
  {Lu}}, \bibinfo {author} {\bibfnamefont {J.~B.}\ \bibnamefont {Surya}},
  \bibinfo {author} {\bibfnamefont {X.}~\bibnamefont {Liu}}, \bibinfo {author}
  {\bibfnamefont {Y.}~\bibnamefont {Xu}}, \ and\ \bibinfo {author}
  {\bibfnamefont {H.~X.}\ \bibnamefont {Tang}},\ }\href {\doibase
  10.1364/OL.44.001492} {\bibfield  {journal} {\bibinfo  {journal} {Opt.
  Lett.}\ }\textbf {\bibinfo {volume} {44}},\ \bibinfo {pages} {1492} (\bibinfo
  {year} {2019})}\BibitemShut {NoStop}%
\bibitem [{\citenamefont {Shoji}\ \emph {et~al.}(1997)\citenamefont {Shoji},
  \citenamefont {Kondo}, \citenamefont {Kitamoto}, \citenamefont {Shirane},\
  and\ \citenamefont {Ito}}]{Shoji97}%
  \BibitemOpen
  \bibfield  {author} {\bibinfo {author} {\bibfnamefont {I.}~\bibnamefont
  {Shoji}}, \bibinfo {author} {\bibfnamefont {T.}~\bibnamefont {Kondo}},
  \bibinfo {author} {\bibfnamefont {A.}~\bibnamefont {Kitamoto}}, \bibinfo
  {author} {\bibfnamefont {M.}~\bibnamefont {Shirane}}, \ and\ \bibinfo
  {author} {\bibfnamefont {R.}~\bibnamefont {Ito}},\ }\href {\doibase
  10.1364/JOSAB.14.002268} {\bibfield  {journal} {\bibinfo  {journal} {J. Opt.
  Soc. Am. B}\ }\textbf {\bibinfo {volume} {14}},\ \bibinfo {pages} {2268}
  (\bibinfo {year} {1997})}\BibitemShut {NoStop}%
\bibitem [{\citenamefont {Zhang}\ \emph {et~al.}(2017)\citenamefont {Zhang},
  \citenamefont {Wang}, \citenamefont {Cheng}, \citenamefont {Shams-Ansari},\
  and\ \citenamefont {Lon\v{c}ar}}]{ZhangOptica}%
  \BibitemOpen
  \bibfield  {author} {\bibinfo {author} {\bibfnamefont {M.}~\bibnamefont
  {Zhang}}, \bibinfo {author} {\bibfnamefont {C.}~\bibnamefont {Wang}},
  \bibinfo {author} {\bibfnamefont {R.}~\bibnamefont {Cheng}}, \bibinfo
  {author} {\bibfnamefont {A.}~\bibnamefont {Shams-Ansari}}, \ and\ \bibinfo
  {author} {\bibfnamefont {M.}~\bibnamefont {Lon\v{c}ar}},\ }\href {\doibase
  10.1364/OPTICA.4.001536} {\bibfield  {journal} {\bibinfo  {journal} {Optica}\
  }\textbf {\bibinfo {volume} {4}},\ \bibinfo {pages} {1536} (\bibinfo {year}
  {2017})}\BibitemShut {NoStop}%
\bibitem [{\citenamefont {Jankowski}\ \emph {et~al.}(2020)\citenamefont
  {Jankowski}, \citenamefont {Langrock}, \citenamefont {Desiatov},
  \citenamefont {Marandi}, \citenamefont {Wang}, \citenamefont {Zhang},
  \citenamefont {Phillips}, \citenamefont {Lon\v{c}ar},\ and\ \citenamefont
  {Fejer}}]{Jankowski20}%
  \BibitemOpen
  \bibfield  {author} {\bibinfo {author} {\bibfnamefont {M.}~\bibnamefont
  {Jankowski}}, \bibinfo {author} {\bibfnamefont {C.}~\bibnamefont {Langrock}},
  \bibinfo {author} {\bibfnamefont {B.}~\bibnamefont {Desiatov}}, \bibinfo
  {author} {\bibfnamefont {A.}~\bibnamefont {Marandi}}, \bibinfo {author}
  {\bibfnamefont {C.}~\bibnamefont {Wang}}, \bibinfo {author} {\bibfnamefont
  {M.}~\bibnamefont {Zhang}}, \bibinfo {author} {\bibfnamefont {C.~R.}\
  \bibnamefont {Phillips}}, \bibinfo {author} {\bibfnamefont {M.}~\bibnamefont
  {Lon\v{c}ar}}, \ and\ \bibinfo {author} {\bibfnamefont {M.~M.}\ \bibnamefont
  {Fejer}},\ }\href {\doibase 10.1364/OPTICA.7.000040} {\bibfield  {journal}
  {\bibinfo  {journal} {Optica}\ }\textbf {\bibinfo {volume} {7}},\ \bibinfo
  {pages} {40} (\bibinfo {year} {2020})}\BibitemShut {NoStop}%
\bibitem [{\citenamefont {Kanashov}\ and\ \citenamefont
  {Rubenchik}(1981)}]{Kanashov}%
  \BibitemOpen
  \bibfield  {author} {\bibinfo {author} {\bibfnamefont {A.}~\bibnamefont
  {Kanashov}}\ and\ \bibinfo {author} {\bibfnamefont {A.}~\bibnamefont
  {Rubenchik}},\ }\href {\doibase https://doi.org/10.1016/0167-2789(81)90009-9}
  {\bibfield  {journal} {\bibinfo  {journal} {Physica D: Nonlinear Phenomena}\
  }\textbf {\bibinfo {volume} {4}},\ \bibinfo {pages} {122 } (\bibinfo {year}
  {1981})}\BibitemShut {NoStop}%
\bibitem [{\citenamefont {Kinsler}\ and\ \citenamefont
  {New}(2003)}]{Kinsler03}%
  \BibitemOpen
  \bibfield  {author} {\bibinfo {author} {\bibfnamefont {P.}~\bibnamefont
  {Kinsler}}\ and\ \bibinfo {author} {\bibfnamefont {G.~H.~C.}\ \bibnamefont
  {New}},\ }\href {\doibase 10.1103/PhysRevA.67.023813} {\bibfield  {journal}
  {\bibinfo  {journal} {Phys. Rev. A}\ }\textbf {\bibinfo {volume} {67}},\
  \bibinfo {pages} {023813} (\bibinfo {year} {2003})}\BibitemShut {NoStop}%
\bibitem [{\citenamefont {Genty}\ \emph {et~al.}(2007)\citenamefont {Genty},
  \citenamefont {Kinsler}, \citenamefont {Kibler},\ and\ \citenamefont
  {Dudley}}]{Genty07}%
  \BibitemOpen
  \bibfield  {author} {\bibinfo {author} {\bibfnamefont {G.}~\bibnamefont
  {Genty}}, \bibinfo {author} {\bibfnamefont {P.}~\bibnamefont {Kinsler}},
  \bibinfo {author} {\bibfnamefont {B.}~\bibnamefont {Kibler}}, \ and\ \bibinfo
  {author} {\bibfnamefont {J.~M.}\ \bibnamefont {Dudley}},\ }\href {\doibase
  10.1364/OE.15.005382} {\bibfield  {journal} {\bibinfo  {journal} {Opt.
  Express}\ }\textbf {\bibinfo {volume} {15}},\ \bibinfo {pages} {5382}
  (\bibinfo {year} {2007})}\BibitemShut {NoStop}%
\bibitem [{\citenamefont {Conforti}\ \emph {et~al.}(2010)\citenamefont
  {Conforti}, \citenamefont {Baronio},\ and\ \citenamefont
  {De~Angelis}}]{Conforti}%
  \BibitemOpen
  \bibfield  {author} {\bibinfo {author} {\bibfnamefont {M.}~\bibnamefont
  {Conforti}}, \bibinfo {author} {\bibfnamefont {F.}~\bibnamefont {Baronio}}, \
  and\ \bibinfo {author} {\bibfnamefont {C.}~\bibnamefont {De~Angelis}},\
  }\href {\doibase 10.1103/PhysRevA.81.053841} {\bibfield  {journal} {\bibinfo
  {journal} {Phys. Rev. A}\ }\textbf {\bibinfo {volume} {81}},\ \bibinfo
  {pages} {053841} (\bibinfo {year} {2010})}\BibitemShut {NoStop}%
\bibitem [{\citenamefont {Wabnitz}\ and\ \citenamefont
  {Kozlov}(2010)}]{Wabnitz10}%
  \BibitemOpen
  \bibfield  {author} {\bibinfo {author} {\bibfnamefont {S.}~\bibnamefont
  {Wabnitz}}\ and\ \bibinfo {author} {\bibfnamefont {V.~V.}\ \bibnamefont
  {Kozlov}},\ }\href {\doibase 10.1364/JOSAB.27.001707} {\bibfield  {journal}
  {\bibinfo  {journal} {J. Opt. Soc. Am. B}\ }\textbf {\bibinfo {volume}
  {27}},\ \bibinfo {pages} {1707} (\bibinfo {year} {2010})}\BibitemShut
  {NoStop}%
\bibitem [{\citenamefont {Baronio}\ \emph {et~al.}(2012)\citenamefont
  {Baronio}, \citenamefont {Conforti}, \citenamefont {Angelis}, \citenamefont
  {Modotto}, \citenamefont {Wabnitz}, \citenamefont {Andreana}, \citenamefont
  {Tonello}, \citenamefont {Leproux},\ and\ \citenamefont
  {Couderc}}]{Baronio12}%
  \BibitemOpen
  \bibfield  {author} {\bibinfo {author} {\bibfnamefont {F.}~\bibnamefont
  {Baronio}}, \bibinfo {author} {\bibfnamefont {M.}~\bibnamefont {Conforti}},
  \bibinfo {author} {\bibfnamefont {C.~D.}\ \bibnamefont {Angelis}}, \bibinfo
  {author} {\bibfnamefont {D.}~\bibnamefont {Modotto}}, \bibinfo {author}
  {\bibfnamefont {S.}~\bibnamefont {Wabnitz}}, \bibinfo {author} {\bibfnamefont
  {M.}~\bibnamefont {Andreana}}, \bibinfo {author} {\bibfnamefont
  {A.}~\bibnamefont {Tonello}}, \bibinfo {author} {\bibfnamefont
  {P.}~\bibnamefont {Leproux}}, \ and\ \bibinfo {author} {\bibfnamefont
  {V.}~\bibnamefont {Couderc}},\ }\href {\doibase
  https://doi.org/10.1016/j.yofte.2012.07.001} {\bibfield  {journal} {\bibinfo
  {journal} {Optical Fiber Technology}\ }\textbf {\bibinfo {volume} {18}},\
  \bibinfo {pages} {283 } (\bibinfo {year} {2012})}\BibitemShut {NoStop}%
\bibitem [{\citenamefont {Hansson}\ \emph {et~al.}(2016)\citenamefont
  {Hansson}, \citenamefont {Leo}, \citenamefont {Erkintalo}, \citenamefont
  {Anthony}, \citenamefont {Coen}, \citenamefont {Ricciardi}, \citenamefont
  {Rosa},\ and\ \citenamefont {Wabnitz}}]{Hansson16}%
  \BibitemOpen
  \bibfield  {author} {\bibinfo {author} {\bibfnamefont {T.}~\bibnamefont
  {Hansson}}, \bibinfo {author} {\bibfnamefont {F.}~\bibnamefont {Leo}},
  \bibinfo {author} {\bibfnamefont {M.}~\bibnamefont {Erkintalo}}, \bibinfo
  {author} {\bibfnamefont {J.}~\bibnamefont {Anthony}}, \bibinfo {author}
  {\bibfnamefont {S.}~\bibnamefont {Coen}}, \bibinfo {author} {\bibfnamefont
  {I.}~\bibnamefont {Ricciardi}}, \bibinfo {author} {\bibfnamefont {M.~D.}\
  \bibnamefont {Rosa}}, \ and\ \bibinfo {author} {\bibfnamefont
  {S.}~\bibnamefont {Wabnitz}},\ }\href {\doibase 10.1364/JOSAB.33.001207}
  {\bibfield  {journal} {\bibinfo  {journal} {J. Opt. Soc. Am. B}\ }\textbf
  {\bibinfo {volume} {33}},\ \bibinfo {pages} {1207} (\bibinfo {year}
  {2016})}\BibitemShut {NoStop}%
\bibitem [{\citenamefont {Bache}(2016)}]{Bache16}%
  \BibitemOpen
  \bibfield  {author} {\bibinfo {author} {\bibfnamefont {M.}~\bibnamefont
  {Bache}},\ }\href@noop {} {\bibfield  {journal} {\bibinfo  {journal}
  {arXiv:{1603.00188}}\ } (\bibinfo {year} {2016})},\ \Eprint
  {http://arxiv.org/abs/1603.00188} {arXiv:1603.00188 [physics.optics]}
  \BibitemShut {NoStop}%
\bibitem [{\citenamefont {Efimov}\ \emph {et~al.}(2005)\citenamefont {Efimov},
  \citenamefont {Yulin}, \citenamefont {Skryabin}, \citenamefont {Knight},
  \citenamefont {Joly}, \citenamefont {Omenetto}, \citenamefont {Taylor},\ and\
  \citenamefont {Russell}}]{Efimov05}%
  \BibitemOpen
  \bibfield  {author} {\bibinfo {author} {\bibfnamefont {A.}~\bibnamefont
  {Efimov}}, \bibinfo {author} {\bibfnamefont {A.~V.}\ \bibnamefont {Yulin}},
  \bibinfo {author} {\bibfnamefont {D.~V.}\ \bibnamefont {Skryabin}}, \bibinfo
  {author} {\bibfnamefont {J.~C.}\ \bibnamefont {Knight}}, \bibinfo {author}
  {\bibfnamefont {N.}~\bibnamefont {Joly}}, \bibinfo {author} {\bibfnamefont
  {F.~G.}\ \bibnamefont {Omenetto}}, \bibinfo {author} {\bibfnamefont {A.~J.}\
  \bibnamefont {Taylor}}, \ and\ \bibinfo {author} {\bibfnamefont
  {P.}~\bibnamefont {Russell}},\ }\href {\doibase
  10.1103/PhysRevLett.95.213902} {\bibfield  {journal} {\bibinfo  {journal}
  {Phys. Rev. Lett.}\ }\textbf {\bibinfo {volume} {95}},\ \bibinfo {pages}
  {213902} (\bibinfo {year} {2005})}\BibitemShut {NoStop}%
\bibitem [{\citenamefont {Dudley}\ \emph {et~al.}(2006)\citenamefont {Dudley},
  \citenamefont {Genty},\ and\ \citenamefont {Coen}}]{Dudley}%
  \BibitemOpen
  \bibfield  {author} {\bibinfo {author} {\bibfnamefont {J.~M.}\ \bibnamefont
  {Dudley}}, \bibinfo {author} {\bibfnamefont {G.}~\bibnamefont {Genty}}, \
  and\ \bibinfo {author} {\bibfnamefont {S.}~\bibnamefont {Coen}},\ }\href
  {\doibase 10.1103/RevModPhys.78.1135} {\bibfield  {journal} {\bibinfo
  {journal} {Rev. Mod. Phys.}\ }\textbf {\bibinfo {volume} {78}},\ \bibinfo
  {pages} {1135} (\bibinfo {year} {2006})}\BibitemShut {NoStop}%
\bibitem [{\citenamefont {Okawachi}\ \emph {et~al.}(2017)\citenamefont
  {Okawachi}, \citenamefont {Yu}, \citenamefont {Cardenas}, \citenamefont {Ji},
  \citenamefont {Lipson},\ and\ \citenamefont {Gaeta}}]{Okawachi17}%
  \BibitemOpen
  \bibfield  {author} {\bibinfo {author} {\bibfnamefont {Y.}~\bibnamefont
  {Okawachi}}, \bibinfo {author} {\bibfnamefont {M.}~\bibnamefont {Yu}},
  \bibinfo {author} {\bibfnamefont {J.}~\bibnamefont {Cardenas}}, \bibinfo
  {author} {\bibfnamefont {X.}~\bibnamefont {Ji}}, \bibinfo {author}
  {\bibfnamefont {M.}~\bibnamefont {Lipson}}, \ and\ \bibinfo {author}
  {\bibfnamefont {A.~L.}\ \bibnamefont {Gaeta}},\ }\href {\doibase
  10.1364/OL.42.004466} {\bibfield  {journal} {\bibinfo  {journal} {Opt.
  Lett.}\ }\textbf {\bibinfo {volume} {42}},\ \bibinfo {pages} {4466} (\bibinfo
  {year} {2017})}\BibitemShut {NoStop}%
\bibitem [{\citenamefont {Webb}\ \emph {et~al.}(2013)\citenamefont {Webb},
  \citenamefont {Xu}, \citenamefont {Erkintalo},\ and\ \citenamefont
  {Murdoch}}]{Webb13}%
  \BibitemOpen
  \bibfield  {author} {\bibinfo {author} {\bibfnamefont {K.~E.}\ \bibnamefont
  {Webb}}, \bibinfo {author} {\bibfnamefont {Y.~Q.}\ \bibnamefont {Xu}},
  \bibinfo {author} {\bibfnamefont {M.}~\bibnamefont {Erkintalo}}, \ and\
  \bibinfo {author} {\bibfnamefont {S.~G.}\ \bibnamefont {Murdoch}},\ }\href
  {\doibase 10.1364/OL.38.000151} {\bibfield  {journal} {\bibinfo  {journal}
  {Opt. Lett.}\ }\textbf {\bibinfo {volume} {38}},\ \bibinfo {pages} {151}
  (\bibinfo {year} {2013})}\BibitemShut {NoStop}%
\bibitem [{\citenamefont {Maddaloni}\ \emph {et~al.}(2016)\citenamefont
  {Maddaloni}, \citenamefont {De~Natale},\ and\ \citenamefont
  {Bellini}}]{maddaloni2016laser}%
  \BibitemOpen
  \bibfield  {author} {\bibinfo {author} {\bibfnamefont {P.}~\bibnamefont
  {Maddaloni}}, \bibinfo {author} {\bibfnamefont {P.}~\bibnamefont
  {De~Natale}}, \ and\ \bibinfo {author} {\bibfnamefont {M.}~\bibnamefont
  {Bellini}},\ }\href@noop {} {\emph {\bibinfo {title} {Laser-based
  measurements for time and frequency domain applications: a handbook}}}\
  (\bibinfo  {publisher} {CRC Press},\ \bibinfo {year} {2016})\BibitemShut
  {NoStop}%
\bibitem [{\citenamefont {He}\ \emph {et~al.}(2019)\citenamefont {He},
  \citenamefont {Zhang}, \citenamefont {Shams-Ansari}, \citenamefont {Zhu},
  \citenamefont {Wang},\ and\ \citenamefont {Lon\v{c}ar}}]{He19}%
  \BibitemOpen
  \bibfield  {author} {\bibinfo {author} {\bibfnamefont {L.}~\bibnamefont
  {He}}, \bibinfo {author} {\bibfnamefont {M.}~\bibnamefont {Zhang}}, \bibinfo
  {author} {\bibfnamefont {A.}~\bibnamefont {Shams-Ansari}}, \bibinfo {author}
  {\bibfnamefont {R.}~\bibnamefont {Zhu}}, \bibinfo {author} {\bibfnamefont
  {C.}~\bibnamefont {Wang}}, \ and\ \bibinfo {author} {\bibfnamefont
  {M.}~\bibnamefont {Lon\v{c}ar}},\ }\href {\doibase 10.1364/OL.44.002314}
  {\bibfield  {journal} {\bibinfo  {journal} {Opt. Lett.}\ }\textbf {\bibinfo
  {volume} {44}},\ \bibinfo {pages} {2314} (\bibinfo {year}
  {2019})}\BibitemShut {NoStop}%
\end{thebibliography}%

\end{document}